# Experimental investigation of an electronegative cylindrical capacitively coupled geometrically asymmetric plasma discharge with an axisymmetric magnetic field

**Swati Dahiya[1, 2], Narayan Sharma[3], Shivani Geete[3], Sarveshwar Sharma[1, 2], Nishant Sirse[*3] and Shantanu Karkari[1, 2]**

[1]*Institute for Plasma Research, Bhat, Gandhinagar, Gujarat, 382428 India*
[2]*Homi Bhabha National Institute, Training School Complex, Anushaktinagar, Mumbai-400094, India*
[3]*Institute of Science and Research and Centre for Scientific and Applied Research, IPS Academy, Indore-452012, India*

Email: nishantsirse@ipsacademy.org

## Abstract

In this study, we have investigated the production of negative ions by mixing electronegative oxygen gas with electropositive argon gas in a geometrically asymmetric cylindrical capacitively coupled radio frequency plasma discharge. The plasma parameters such as density (electron, positive and negative ion), negative ion fraction, and electron temperature are investigated for fixed gas pressure and increasing axial magnetic field strength. The axisymmetric magnetic field creates an E×B drift in the azimuthal direction, leading to the confinement of high-energy electrons at the radial edge of the chamber, resulting in decreased species density and negative ion fraction in the plasma bulk. However, the electron temperature increases with the magnetic field. It is concluded that low magnetic fields are better suited for negative ion production in such devices. Furthermore, in addition to the percentage ratio of the two gases, the applied axial magnetic field also plays a vital role in controlling negative ion fraction.

## I.   INTRODUCTION

Stable electropositive argon plasma mixed with a small amount of reactive gas such as oxygen has a wide range of applications in the semiconductor industry for processes like silicon wafer etching[1–5], thin film deposition[6–8], negative ion sources[9–12], formation of oxide film[13,14], etc.  The admixture of reactive gases not only provides necessary radicals for surface processes but also influences the plasma parameters. The control of parameters like density and temperature is essential for optimizing the deposition and etching process to get the film of the desired structure and morphology. Therefore, studying the nature and properties of plasma formed by the mixture of electropositive and electronegative gas is an important area of research. All these applications are influenced by the negative ion concentration in the discharge. Therefore, it is worthwhile to investigate the plasma parameters and the negative ion density in such discharges. Various authors studied negative ions production in argon-oxygen discharge. Shindo *et al.* measured the negative ion density by using a Langmuir probe and ion acoustic wave (IAW) method in the Argon + Oxygen plasma in an Electron Cyclotron Resonance (ECR) discharge chamber and have measured the negative ion fraction ($\alpha = n_-/n_e$) of the order of 0.3. They have also reported that α is strongly depends on the gas mixture ratio and has the optimum value when the oxygen percentage in the discharge is about 15 percent[15]. Katsch *et al.* studied the negative ion fraction in pulsed inductively excited plasma. They have found that a high negative ion fraction occurs at low pressure and high input power[16]. Gudmundsson *et al.* experimentally investigated the plasma parameters in the argon-oxygen plasma and compared the measured parameters with the values obtained by using the volume-averaged global model assuming Maxwellian electron energy distribution function (EEDF)[17]. Gudmundsson *et al.* also employed a volume-averaged global model to study the dissociation process of oxygen molecules in the argon-oxygen discharge. They found that the dissociation fraction of the oxygen molecule increases with the increase in the argon content in the discharge[18]. Kitajima *et al.* reported the increased growth rate of $SiO_2$ when stable gas like argon or krypton is mixed with a small amount of oxygen. They observed a two-fold increase in the oxygen metastable state density with the dilution of the argon gas[4]. Other notable works related to the argon plasma diluted with oxygen gas are done by Takechi *et al*[1], Lee *et al.*[19], Wang *et al.*[20], Bradley *et al*[21], Worsley *et al*[5], etc.

It is well known that low-pressure RF discharge produces higher plasma density at lower RF power as compared to DC discharges. Therefore, capacitively coupled low-pressure RF discharges have various applications in microelectronic fabrication in solar cells for etching and deposition of thin films[22–26], semiconductor industry[27–31], biomedical applications[32–34], space technology[32,33], etc. Generally, molecular gas mixed with novel gas is essential in optimizing the etching and deposition rate in film deposition and plasma processing. A mixture of two or more gases is used in the plasma processing chamber to achieve the desired morphology of the deposited film. In order to improve the efficiency of such processes, a good understanding of the plasma system is necessary. A simple CCP discharge system consists of two parallel plates, in which one is RF-powered, and the other is grounded. Substrate is generally placed in the powered electrode for plasma processing[22,23]. Various works have been done to enhance the efficiency of such devices by independently controlling the flux and energy of the impinging ions. Some of the techniques adopted are using dual or multiple frequency RF sources[35–44], tailored current and voltage waveforms[45–52], and operating the system in the pulsed power mode[53–55]. It is observed that higher plasma density is produced by applying the higher RF frequency which increases the ion flux to the substrate. However, with the increased driving frequency, the plasma uniformity decreases, reducing such system's efficiency. To overcome this problem, Hervey *et al.* used multi-tile electrodes at very high frequencies[56], Chabert *et al.* used shaped electrode and dielectric lens[57], Schmidt *et al.* used a lens-shaped circular electrode to improve the plasma uniformity[58]. The performance of a capacitively coupled plasma source can also be enhanced by applying the static magnetic field perpendicular to the static electric field. The applied magnetic field reduces the radial loss of plasma to the wall thereby enhancing the density. Apart from enhancing the performance of the CCP, the applied magnetic field provides the interaction mechanism between the RF field and plasma. Hence, the external magnetic field provides additional scope to control the properties of the discharge in order to make it suitable for particular applications[59–62]. In magnetically enhanced reactive ion etching (MERIE), the transverse magnetic field is applied to the capacitively coupled plasma sources to increase the plasma density by reducing the plasma loss across the field lines[61]. You *et al.*[63] and Lee *et al.*[64] studied the transition of electron kinetic properties from non-local to local by applying a static magnetic field to the CCP discharges. Thus, the magnetic field acts as the controlling parameter to tune the plasma parameters like plasma density and temperature in the discharge to create a region of uniform ion flux suitable for plasma processing. It implies that applying a static magnetic field in the CCP discharges can control the power transfer mechanism. Trieschmann *et al*[65] and Yang *et al*.[66,67] demonstrated that the application of an asymmetric magnetic field to the capacitively coupled discharge, DC self-bias as well as asymmetric plasma response can be activated even in the geometrically and electrically symmetric system. This phenomenon is known as the magnetic asymmetric effect. This effect can independently control the ion flux and energy. Thus, in the presence of a transverse magnetic field, many physical processes like E×B drift, axial variation of plasma parameters, mode transition, and power dissipation can occur, influencing the performance and efficiency of the CCP discharge source for industrial applications.

However, the E×B drift in the case of a parallel plate CCP generates the non- uniform plasma density and electron temperature profile[68]. This configuration affects the performance of parallel plate CCP discharge for industrial applications. Cylindrical electrodes CCP having axisymmetric magnetic fields can increase the plasma density and uniformity by radially confining the electrons. Such a system can be designed using two co-axial cylinders, while axisymmetric magnetic field can be generated using electromagnets[69,70].

Considering the applications of stable electropositive argon plasma diluted with a small amount of reactive electronegative gas such as oxygen in the microelectronics and semiconductor industry, and the advantages of cylindrical electrodes CCP with axisymmetric magnetic field, in the present study we have studied the negative ion production in Argon + Oxygen discharge in geometrically asymmetric cylindrical capacitively coupled RF plasma source. By varying the magnetic field strength from 0-8 millitesla (mT), we have studied the variation of plasma parameters like density (electron, positive and negative ion), negative ion fraction as well as electron temperature with the changes of discharge parameters like applied RF power and axial magnetic field. In this experiment, we have used RF compensated Langmuir probe to measure the negative ion fraction in the bulk plasma. Langmuir probe has various advantages. It can provide spatially resolved data and reach to the remote locations inside the electrode. Moreover, Langmuir probe is a cost-effective diagnostic that can be easily fabricated in-house.

In this work, we have investigated the possibility of using CCP cylindrical electrodes plasma source with an axisymmetric magnetic field for the production of negative ion density argon oxygen mixed plasma. The paper is structured as follows: Section 2, describes the cylindrical electrodes CCP plasma source with an axisymmetric magnetic field and the Langmuir probe diagnostics technique adopted for measuring negative ion fraction. Section 3 is devoted to the results and discussion. Finally, the paper is concluded with the summarization and the outcome of the studies in section 4.

## II. EXPERIMENTAL SET UP AND DIAGNOSTICS

### A. EXPERIMENTAL SET UP

Fig 1 shows the schematic of the plasma chamber, diagnostic setup and the electrode assembly. The system consists of a large vacuum chamber of 31 cm diameter and 120 cm length. The chamber is evacuated using a Pfeiffer make 700 l/s turbo molecular pump (TMP) backed by a rotary pump up to the base pressure of $2\times10^{-5}$ mbar. Inside the vacuum chamber, a Cylindrical Electrode (CE) of 20 cm in length and an inner diameter of 24 cm is placed, which is powered by a 13.56 MHz RF power supply and matching network. At both ends of the powered electrode, there are two annular rings of 19 cm inner and 23 cm outer diameter. Annular rings are followed by two grids, $G_1$ and $G_2$ on both ends of the powered electrodes. Grids $G_1$ and $G_2$ are electrically isolated using insulator blocks. Grid $G_1$ is grounded and is a fine grid, whereas Grid $G_2$ is coarse and is electrically floated. The whole electrode assembly is isolated using the ceramic blocks. An Axial magnetic field is produced by using two electromagnets. The two electromagnets are kept at 27 cm apart from each other. The electromagnets are powered by a DC power supply having a maximum current rating of 60 A. A more detailed description of the set-up, and the variation of the axial magnetic field with the changes in the value of direct current applied to the electromagnet is given in the references[69,70].

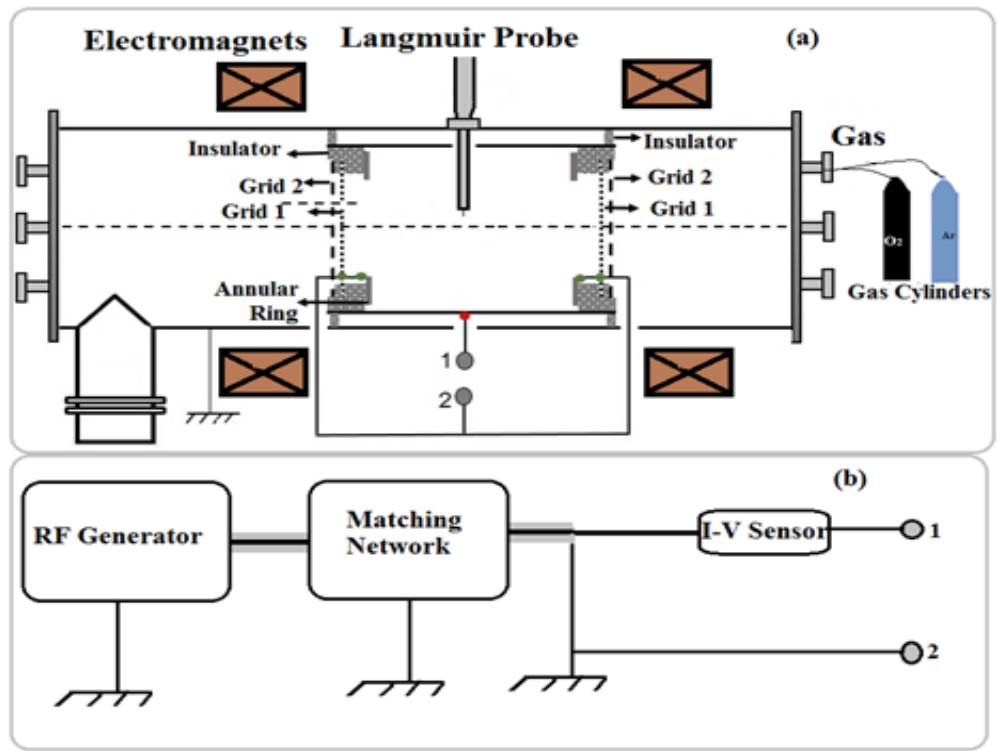

Fig.1. (a) Schematic of the plasma chamber showing the electrode assembly. (b) The discharge circuit for powering the electrode.

The explained electrode assembly is kept in the uniform axial magnetic field region. An RF-compensated Langmuir probe is inserted radially at the axis of the chamber through the small hole, cut at the centre of the cylindrical electrode. Pure argon plasma and argon oxygen mixed plasma with varying oxygen percentages from 10 percent to 30 produced inside the cylindrical electrode at the working pressure of $1\times10^{-3}$ mbar. The density variation of different species (electron, positive and negative ion density) as well as electron temperature and negative ion fraction with the applied magnetic field for different values of RF power at the different concentrations of oxygen gas is investigated in this work.

## B. Diagnostics

Plasma parameters are measured using an RF-compensated Langmuir probe described in our previous works[69-70]. The effect of fundamental frequency (13.56 MHz) and its first harmonics (27.12 MHz) is compensated by using a resonant filtering technique. In this method, the resonant circuit is made by connecting self-resonating inductors to the probe tip, which provides high impedance for fundamental (180 kΩ) and first harmonic (25 kΩ). The main probe tip is made up of tungsten wire of diameter 0.2 mm and 5 mm length. A floating auxiliary probe with area greater than the main probe tip is connected to the main probe by using a 1 *nF* capacitor. The auxiliary probe reduces input RF impedance across the probe sheath. I-V characteristics are obtained under different operating conditions by driving the main probe with an amplified voltage sweep from high voltage amplifier (WMA-300, Falco Systems) and a unity gain isolation amplifier circuit using AD215 isolation amplifier. Orbital Motion Limited (OML) theory is used to calculate the plasma density. In the present study, the magnetic field strength is varied from 0-8 mT, which correspond to electron gyro radius from ~0.05–0.4 cm and is larger than the probe radius (0.01 cm). Additionally, the Langmuir probe is inserted across the field lines to reduce the influence of field lines on the plasma by the magnetic field. Therefore, the influence of magnetic field on measured I-V is minimal and a non-magnetized approximations are valid for the estimation of plasma parameters. The electron temperature is calculated from the logarithmic slope of I-V curve. Plasma potential is derived from either the first or double derivative of the I-V curve.

For the measurement of negative ion density direct technique such as pulse laser photodetachment are oftenly used. However, not possible in the current set-up due to limited accessibility. Thus, we adopted the method of comparing the ratio of positive to negative ions saturation current (as shown in equation 1) of pure argon and argon/oxygen mixed plasma to calculate the negative to positive ion density ratio ($\beta$) [71–73]. Shindo *et al*. measured the negative ion fraction by using the current method shows that $n_-/n_+$ measured by using other techniques shows comparable values upto $n_-/n_+$ ratio of 0.6[71-75]. When the ratio becomes greater than 0.6, then the influence of the positive ion sheath due to the presence of negative ion in the plasma becomes appreciable. In such case, a correction coefficient to account for the sheath effect should be introduced in equation (1). However, in our case, the value of $n_-/n_+ < 0.6$, therefore we have used the equation without any correction factor.

$$\beta = \frac{n_-(X)}{n_+(X)} = 1 - \frac{I_+(Ar)I_{es}(X)}{I_+(X)I_{es}(Ar)}\sqrt{\frac{M_+(Ar)}{M_+(X)}} \qquad (1)$$

Where, $I_+(Ar)$ and $I_+(X)$ are positive ion saturation currents of pure argon and argon-oxygen mixed plasma, $I_{es}(Ar)$ and $I_{es}(X)$ are electron saturation current of pure argon and argon-oxygen mixed plasma and $M_+(Ar)/M_+(X)$ is the ratio of positive ion mass of pure argon and argon-oxygen mixed plasma respectively. All these ratios are determined from the I-V characteristics of the Langmuir probe. $M_+(X)$ is the mean ion mass in the case of argon-oxygen mixed plasma and is calculated using the procedure mentioned in the references[73–75]. The positive ion current ($I_+$) collected by the cylindrical probe at the voltage V can be approximated as

$$I_+(X) \approx eSn_+(X)\sqrt{\frac{eV}{M_+(X)}} \quad \text{or} \quad M_+(X) \approx \left[\frac{eSn_+(X)}{I_+(X)}\right]^2 \qquad (2)$$

Where $n_+(X)$ is the total density of positive ions, $S$ is the surface area of the probe and $M_+(X)$ is the mean ion mass in the case of argon-oxygen mixed plasma. If $h_{pr} = n_+(X)/n_e(X)$, or $n_+(X) = h_{pr}n_e$, then

$$M_+(X) = \left[\frac{eSn_e h_{pr}}{I_+(X)}\right]^2 \qquad (3)$$

Measured values of $n_e$ (Ar) and $I_+$ (V) at V=40 V to satisfy $M_+$ (Ar) = 40 in the case of pure argon plasma are used to calculate the value of the term $h_{pr}$. By approximating this value of $h_{pr}$, the values of $M_+(X)$ for different percentages of argon–oxygen mixture are calculated. It is to be noted that for the calculation of $M_+(X)$, the value of $h_{pr}(X)$ is approximated based on the value of $h_{pr}$ (Ar). It is reasonable to assume that addition of small quantity of oxygen (10% to 30%) to the argon discharge does not affect much on the values of $M_+(X)$. Additionally, the major positive ion of the oxygen discharge (i.e. $O_2^+$) has the mass almost comparable to that of argon positive ion $Ar^+$. By substituting the value of $M_+(X)$ in equation (1), the ratio of the negative ion density to the positive ion density ($\beta$) is calculated. As discussed above, the sheath effect has negligible influence on the value of plasma parameters and hence on $M_+(X)$.

The calculated value of $\beta$ along with the quasineutrality condition can be utilized to calculate the negative ion density and negative ion fraction ($\alpha=n_-/n_e$) as

$$n_+(X) = \frac{n_e(X)}{1-\beta(X)} \qquad (4)$$

$$n_-(X) = n_+(X)\beta(X) \qquad (5)$$

### III.  RESULTS AND DISCUSSION

The chamber is evacuated up to the base pressure of $\sim 2\times 10^{-5}$ mbar using a turbo molecular pump backed by a rotary pump. All the experiments reported in this work are performed at the $1\times 10^{-3}$ mbar working pressure. At first, pure argon is injected into the chamber, then subsequently 90 percent argon and 10 percent oxygen, 80 percent argon and 20 percent oxygen, and finally 70 percent argon and 30 percent oxygen are injected into the chamber. The error bars for the plasma parameters are calculated by repeating the experiment at same discharge conditions. The errors in the electron temperature are then determined by taking multiple sets of linear fit of the semi-log plot, followed by a standard deviation for all the values. Similarly, the technique is repeated for the ion saturation current for the error estimation in the density values.

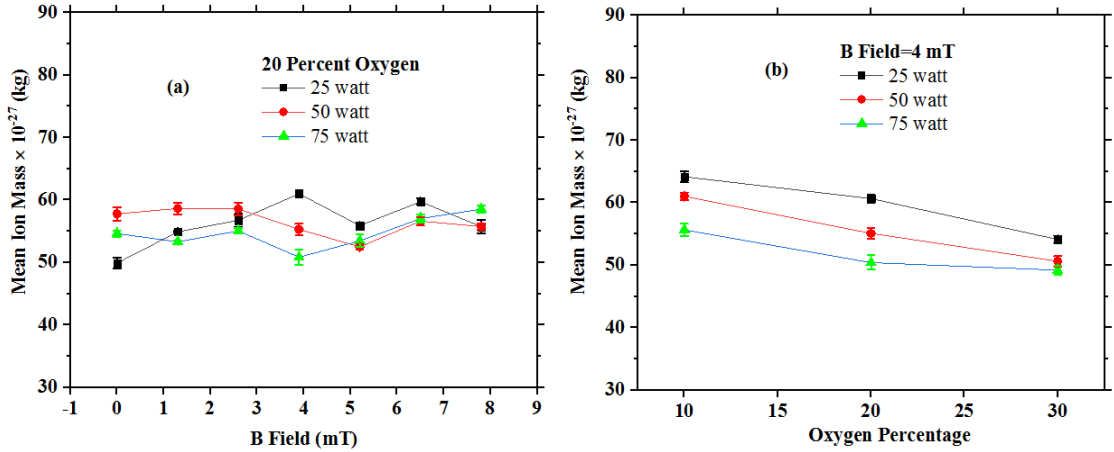

Fig.2. Variation of mean positive ion mass ($M_+(X)$) in the argon-oxygen mixed plasma for (a) 20 percent of oxygen for 25 watt, 50 watt and 75 watt with the increasing applied axisymmetric magnetic fields of the electromagnet and (b) with the increasing percentage of oxygen gas for the applied RF power of 25 watt, 50 watt and 75 watt.

Fig.2 (a) shows the variation of mean positive ion mass ($M_+(X)$) calculated by using equation (3) in argon-oxygen mixed plasma at 20 percent oxygen concentration in the discharge, and for different values of applied RF power. It is observed that the mean ion mass lies approximately between 45-55 × $10^{-27}$ kg. Since the dominant positive ions present in the discharge are $Ar^+$, $O^+$, and $O_2^+$ ions. Thus, this method estimates the reasonable values of positive ions present in the discharge. Fig. 2 (b) shows the variation of the mean ion mass with the increasing oxygen concentration from 10 percent to 30 percent in the argon-oxygen discharge at the externally applied axial magnetic field of 4 mT. It is observed that mean ion mass decreases slowly and steadily with the increasing oxygen concentration in the discharge. Since with the increasing oxygen gas, the concentration of $O^+$ and $O_2^+$ increase and the concentration of $Ar^+$ decreases, and $Ar^+$ ion is heavier as compared to $O^+$ and $O_2^+$ ions, the mean ion mass remains approximately similar trend with the externally applied magnetic field at the given RF power and oxygen concentration whereas decreases steadily with the increasing oxygen gas concentration. However, a small fluctuation in the value of mean ion mass is observed, which may be due to the application of RF power to the electrodes and also may be due to the limitation of diagnostics. This calculated value of mean ion mass is substituted in equation (1) for the calculation of negative to positive ion density ratio ($β$).

Fig 3 (a) and (b), respectively shows the variation of electron density and electron temperature with the applied axial magnetic field for 10 percent oxygen in the discharge. For a low (25 to 75 watt) operating power, the measured plasma density is higher of the order of $10^{16}$ m$^{-3}$ due to enhanced power coupling efficiency in RF discharge. For an unmagnetized case (B=0), the plasma density varies from ~8.5×$10^{15}$ m$^{-3}$ to ~1.3×$10^{16}$ m$^{-3}$, with ~50 % rise, when the applied RF power to the electrode is increased from 25 watt to 75 watt. As magnetic field strength increases, the electron density shows transition from higher value (below 4 mT) to lower value (above 4 mT). At low RF powers (25 watt and 50 watt), the transition is sudden, whereas smooth transition is observed at 75 watt RF power.

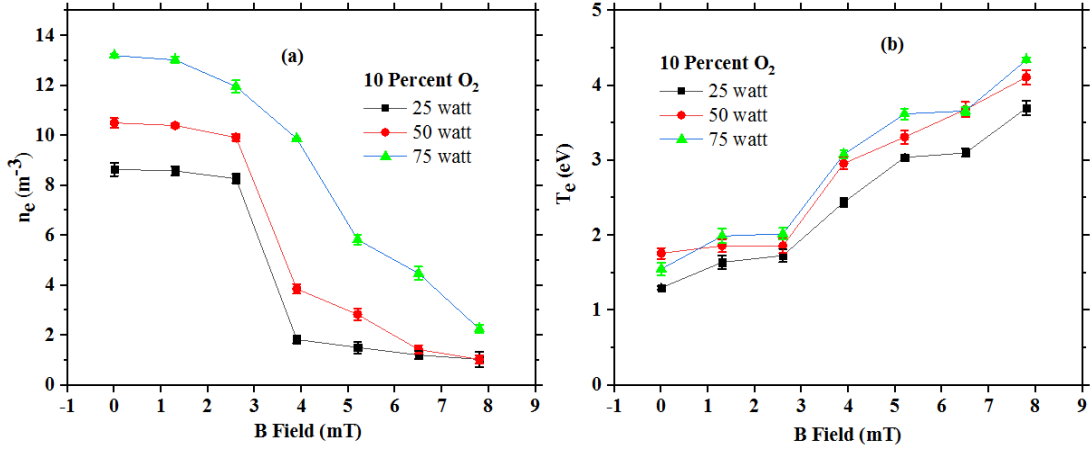

Fig. 3. Variation of (a) electron density and (b) electron temperature for 10 percent oxygen with the increasing values of applied axial magnetic field in the argon-oxygen mixed plasma at the applied RF power of 25 watt, 50 watt and 75 watt.

The variation of electron temperature with the magnetic field shows opposite trends i.e., in low magnetic field regime (below 4 mT), the electron temperature is low and nearly constant (1-2 eV); however, as the magnetic field increases to 4 mT, the $T_e$ increases from 2.5-3 eV, thereafter upon increasing the magnetic field, the electron temperature increases steadily. The opposite trend followed by the density and temperature can be attributed to the power balance equation[22,76]. The particle balance equation in the case of mixed plasma can be written as:

$$n_e = \frac{P_{abs}}{e\mu_B A_{eff}\left(\varepsilon_e + \varepsilon_i + \varepsilon_c^{O_2}\frac{n_{O_2^+}}{n_e} + \varepsilon_c^{O}\frac{n_{O^+}}{n_e} + \varepsilon_c^{Ar}\frac{n_{Ar^+}}{n_e}\right)} \quad (6)$$

Here, $P_{abs}$ is the power absorbed, $\mu_B$ is the Bohm velocity given by $\mu_B = \sqrt{\frac{KT_e}{m_+(X)}}$, $A_{eff}$ is the effective area for the particle loss, $A_{eff} = \frac{V}{d_{eff}}$, expressed as ratio of the volume of the chamber (V) and the effective plasma length ($d_{eff}$). $\varepsilon_T$ is the total energy lost per electron-ion pair from the system and is expressed as the sum of $\varepsilon_T = \varepsilon_e + \varepsilon_i + \varepsilon_c$ energy lost per electron ($\varepsilon_e$), the energy lost per ion ($\varepsilon_i$), and the collisional energy lost per electron-ion pair ($\varepsilon_c$). In the case of oxygen & argon gas mixture, the dominant positive ions present in the plasma are $O_2^+$, $O^+$ and $Ar^+$. Therefore, in the case of mixed plasma, $\varepsilon_c$ is determined using the following equation:

$$\varepsilon_c = \varepsilon_c^{O_2}\frac{n_{O_2^+}}{n_e} + \varepsilon_c^{O}\frac{n_{O^+}}{n_e} + \varepsilon_c^{Ar}\frac{n_{Ar^+}}{n_e} \quad (7)$$

In the power balance equation, density varies inversely with the electron temperature, for the given values of absorbed power ($P_{abs}$) and effective area for the particle loss ($A_{eff}$). Therefore, as the plasma density increases, the electron temperature decreases, as shown in the Fig. 3.

A transition in the electron density and temperature is related to the Lamour radius compare to the sheath thickness. Fig. 4 shows a plot of variation in the sheath thickness and Larmor radius with the increasing value of the applied magnetic field strength for 10 percent oxygen at 75 watt applied RF power. In this plot, Child-Langmuir formula $R_{sh} = \frac{\sqrt{2}\lambda_{Ds}}{3}\left(\frac{2V_0}{T_e}\right)^{\frac{3}{2}}$ is used to calculate the sheath thickness[22]. Here, $\lambda_{Ds}$ is the Debye length and $V_0$ is the electrode voltage measured by using commercial I-V sensor (Impedans Ltd) as shown in Fig. 1.

At no magnetic field, the Larmor radius is infinite, so the electrons produced near the sheath are free to diffuse axially. Therefore, the central plasma density is higher. At 1-3 mT of magnetic field, the Larmor radius is higher than the sheath thickness; hence the peripheral electrons diffuse to the axial region of the discharge tube thereby increasing the electron density at the axis of the cylindrical electrode. However, when the magnetic field increases beyond 4 mT, the Larmor radius decreases and becomes lower than the sheath thickness. As a result, the higher energy electrons are confined near the cylinder's radial edge[70], decreasing the central plasma density, as shown in Figs. 3 and 5. An approximately similar variation of Larmor radius and sheath thickness is observed for other values of magnetic field and oxygen concentration.

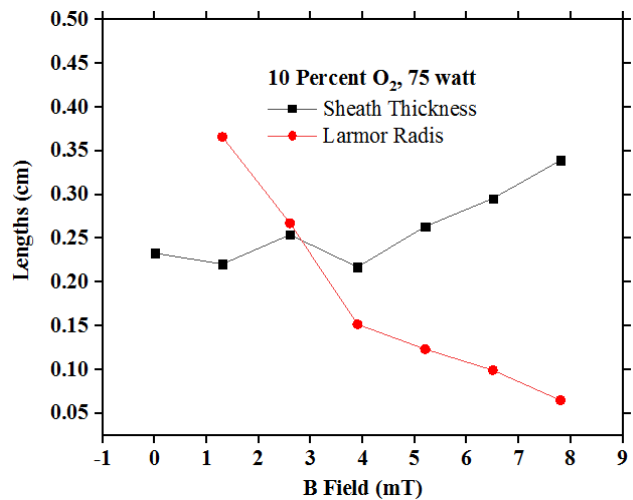

Fig.4. Plot of the variation of the sheath thickness and Larmor radius with the increasing magnetic field for 10 percent oxygen at 75 watt of applied RF power.

In the present discharge system, electric field is in the radial direction and magnetic field in the axial direction, therefore E×B drift is produced in the azimuthal direction with the drift velocity given by $v_{BE} = E/B$. The closed E×B drift enhances ionization at the radial edge due to collision. However, with the increase of magnetic field at the given power, the drift velocity ($v_{BE}$) and hence the enhanced ionization due to collision decrease. This is another factor responsible for the decrease in plasma density with the increase of magnetic field at a fixed RF power.

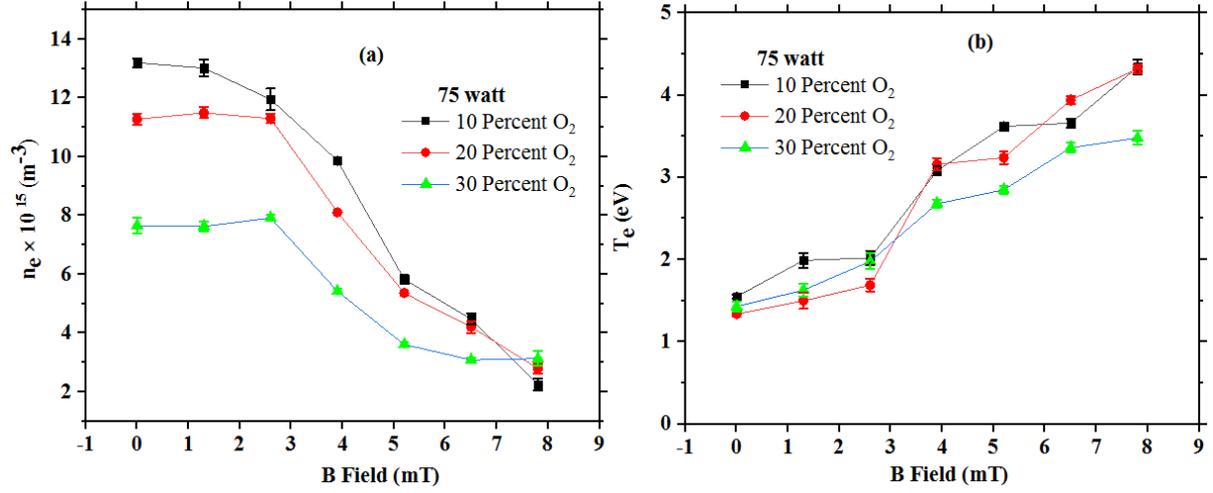

Fig. 5. Variation of (a) plasma density and (b) electron Temperature with the increasing values of applied axial magnetic field for 75 watt RF power for 10 percent, 20 percent, and 30 percent oxygen in the discharge.

Fig. 5 (a) shows the variation of plasma density with the axial magnetic field for 75 watt RF power, and 10 percent, 20 percent, and 30 percent oxygen concentration in the discharge. Plasma density is low for higher oxygen percentage in the discharge. In equation (6), with the increasing oxygen percentage in the discharge, $d_{eff}$ decreases, which leads to an increase in the value of $A_{eff}$[76]. The value of $\varepsilon_e$ and $\varepsilon_i$ is very low compared to $\varepsilon_c$[17]. The expression for $\varepsilon_c$ is given by the equation [17,18].

$$\varepsilon_c = \varepsilon_{iz} + \sum_i \varepsilon_{ex,i} \frac{K_{ex,i}}{K_{iz}} + \frac{K_{el}}{K_{iz}} \frac{3m_e}{m_i(X)} T_e \qquad (8)$$

Where $\varepsilon_{iz}$ is the ionization energy, $\varepsilon_{ex,i}$ is the excitation energy for the i[th] excitation process, $K_{ex}$, $K_{iz}$, and $K_{el}$ are the excitation, ionization and elastic scattering rate constant. By considering ionization ($\varepsilon_{iz}$) and excitation energy ($\varepsilon_{iz}$) as well as elastic scattering ($K_{el}$), ionization ($K_{iz}$), and excitation ($K_{ex}$), the rate constant for the reactions of Ar, O, and $O_2$ as mentioned in the paper by Gudmundsson *et al.* and the reference therein [17], we can plot $\varepsilon_c$ as a function of $T_e$ as shown in Fig. 6.

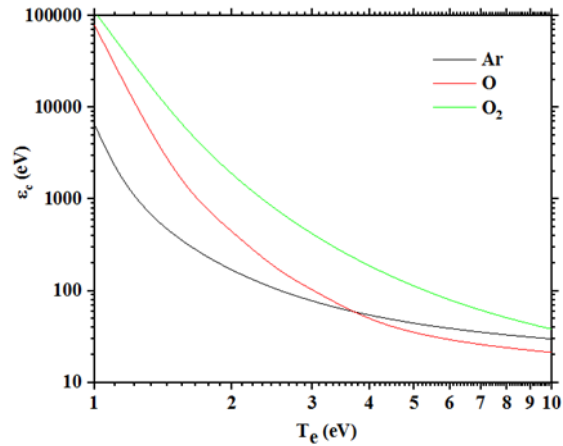

Fig.6. Plot of collisional energy ($\varepsilon_c$) lost per electron-ion pair created as the function of electron temperature for Ar, O, and $O_2$.

For 1 to ~ 4 eV electron temperatures, collisional energy ($\varepsilon_c$) lost per electron-ion pair created for Ar is less than O, and $\varepsilon_c$ of for $O_2$ is higher as compared to that of O and Ar as shown in Fig. 6. Oxygen gas primarily exists in molecular form. For 10, 20, and 30 percent oxygen, argon is the dominant gas in the discharge. A slight increase in oxygen gas in argon discharge increases the values of $A_{eff}$ and $\varepsilon_T$; therefore, as per the power balance equation (equation (6)), there is a small decrease in electron density. In other words, we can say that as oxygen is a molecular gas, higher energy is required for ionization as compared to the argon gas. Therefore, electron density decreases as the oxygen concentration is increased in the argon discharge.

Fig. 5 (b) shows the variation of plasma temperature with the axial magnetic field at 75 watt RF power at 10 percent, 20 percent, and 30 percent oxygen in the discharge. The electron temperature lies between 1-2 eV for the applied magnetic field from 0-3 mT, but as the magnetic field is increased above 4 mT, the temperature increases to 3-4 eV. For 10 and 20 percent oxygen in the discharge, the variation of electron temperature with the applied magnetic field is comparable. Since the electron temperature is measured at the axis of the chamber near the inside of the cylindrical electrode, the applied RF influences it; therefore, some small fluctuation in the electron temperature is observed. The fluctuation of the electron temperature may be also possibly due to the limitation of Langmuir probe diagnostics. However, at 30 percent oxygen, the electron temperature is relatively stable and is slightly lower than at 10 and 20 percent oxygen. According to the particle balance model, by substituting the value of $\mu_B$, we get, $KT_e = K_{iz}^2 d_{eff}^2 m_+(X)$. The ionization rate constant $K_{iz}$ for Ar, O, and $O_2$ is approximately similar for electron temperature between 1-3 eV [73]. The effective plasma length $d_{eff}$ decreases with the increasing oxygen concentration[76]. Also, mean positive ion mass ($M_+(X)$) decreases slightly with the oxygen concentration. Due to the slight decrease of $d_{eff}$ and $M_+(X)$ with the increasing oxygen concentration, the value of electron temperature decreases with the oxygen concentration in the discharge.

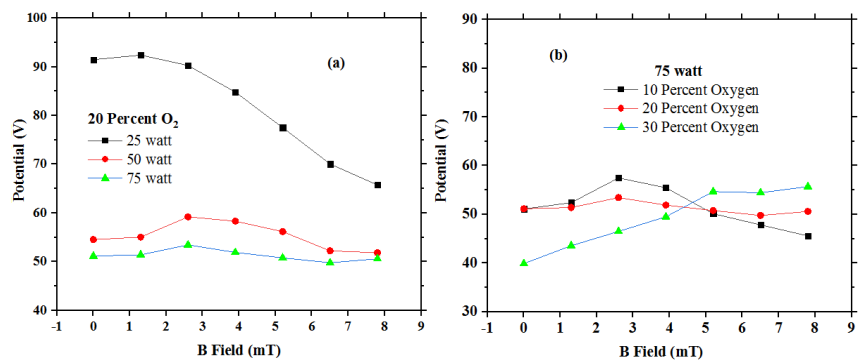

Fig.7. Plot of the plasma potential as the function of the axial magnetic field (a) for 20 percent oxygen at 25 watt, 50 watt, and 75 watt, and (b) at 75 watt for 10 percent, 20 percent and 30 percent oxygen in the discharge.

Fig. 7 (a) shows the plot of the variation of plasma potential with the axial magnetic field for 20 percent of oxygen in the discharge. It is observed that the plasma potential is very high when the applied RF power is 25 watt as compared to 50 watt and 75 watt, and the potential decreases moderately with the axially applied magnetic field. There is a higher loss of electrons to the walls at lower RF power due to lower electron density and/or higher temperature. In order to reduce this loss and to maintain the quasineutrality in the plasma volume, the plasma potential is higher. However, when the axial magnetic field is increased, the plasma losses to the wall decrease due to the radial confinement of electrons. As a result, the plasma potential decreases gradually with the axial magnetic field. However, for 50 and 75 watt RF power decrease in the plasma potential with the axial magnetic field is not substantial, when compared to 25 watt RF power. Fig.7 (b) shows the variation of plasma potential as a function of the axial magnetic field at 75 watt applied RF power for 10 percent, 20 percent, and 30 percent oxygen in the discharge. An interesting observation is that, at the lower value of the axial magnetic field below 4 mT, the plasma potential for 10 percent oxygen is highest and decreases subsequently for 20 and 30 percent oxygen in the discharge. However, at the magnetic field above 4 mT the trend completely reverses. The variation of the trend of plasma potential with the changes in the oxygen concentration in the discharge is due to changes in the composition of the ions i.e., $Ar^+$, $O^+$, $O_2^+$ escaping to the reactor's walls, as well as due to the changes in the values of negative ion fraction with the changes in the oxygen percentage.

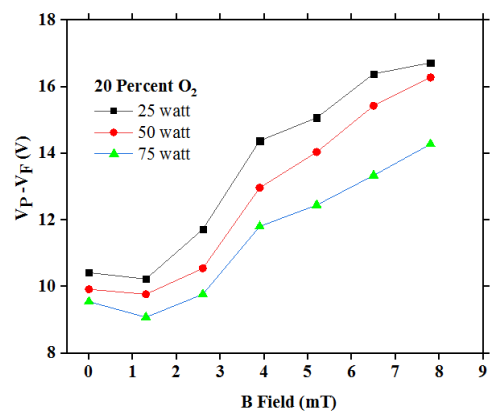

Fig.8. Plot of difference in plasma potential and floating potential ($V_P$-$V_F$) versus magnetic field strength for 20 percent oxygen and 25 watt, 50 watt, and 75 watt.

Fig. 8. shows the variation of difference between plasma potential and floating potential ($V_P$-$V_F$) as the function of applied magnetic field strength when varied from 0-8 mT for the applied power of 20 percent oxygen at 25 watt, 50 watt, and 75 watt. $V_P$-$V_F$ follows approximately the similar trends as the electron temperature (Fig.3 (b)) i.e., lower in the magnetic field regime below 4 mT and increasing thereafter. This is due to the fact that $V_P$-$V_F$ is proportional to the electron temperature., however, in highly electronegative plasma, there is considerable deviation of $V_P$-$V_F$ from the electron temperature. But in our case, the plasma is weakly electronegative due to lower percentage of oxygen in the discharge and thus $V_P$-$V_F$ follows the same trend as electron temperature. These results shows that the modification of sheath due to the presence of negative ions will have negligible influence on the estimated plasma parameters.

Fig. 9 (a-c) shows the plot of the variation of positive ion, electron, and negative ion density with the axial magnetic field at 75 watt of applied RF power when the oxygen concentration in the argon discharge is varied from 10 percent to 30 percent. In all three cases, the variation of the densities with the axial magnetic field follows similar trends. The electron density closely follows positive ion density, whereas negative ion density is much lower. In the lower magnetic field regime (below 4 mT), the gap between the positive ion density and electron density is more, which shows that the negative ion density is higher to maintain quasineutrality. This is due to lower electron temperature in this range (Fig. 3(b) & Fig.5 (b)) as the negative ions are generally formed by the dissociative attachment of low energy electrons to the metastable oxygen molecule i.e. $e + O_2^M \rightarrow O^- + O$. But for the magnetic field higher than 4 mT, the electron density comes very close to the plasma density, indicating the lower negative ions in this region. However, at higher oxygen concentrations in the discharge, the density of all the charged particles is lower; the reason for this effect is already explained above. Therefore, to estimate the qualitative variation of negative ions in the discharge with the oxygen concentration, it is necessary to study the variation of negative ion fraction ($\alpha = n_-/n_e$).

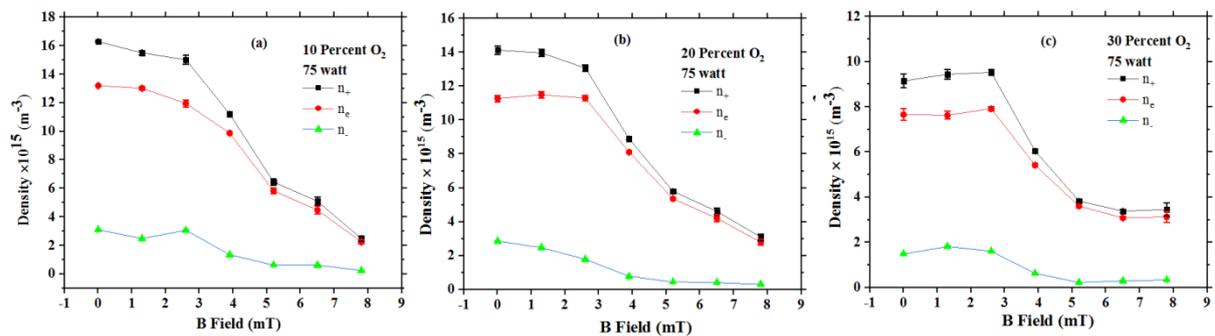

Fig. 8. Plot of the variation of positive ion, electron and negative ion density with the axial magnetic field at the 75 watt of applied RF power when the oxygen concentration in the argon discharge is (a) 10 percent, (b) 20 percent, and (c) 30 percent.

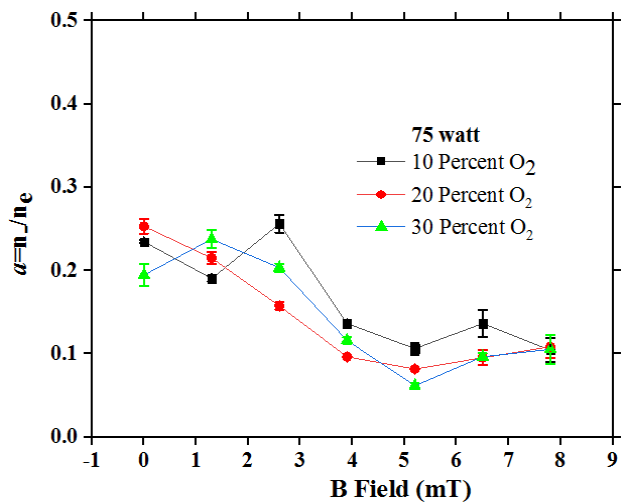

Fig. 9. Plot of the variation of negative ion fraction with the axial magnetic field at the 75 watt of applied RF power when the oxygen concentration in the argon discharge is 10 percent, 20 percent and 30 percent.

Fig. 9. shows the variation of negative ion fraction (α) with the axial magnetic field at 75 watt of applied RF power when the oxygen concentration in the discharge is varied from 10 percent to 30 percent. The negative ion fraction is higher at the lower values of the applied

axial magnetic field, as can be seen from Fig. 9. However, we have observed that the slightly higher negative ion fraction at the lower concentration of oxygen gas (10 percent) for some magnetic field strength. This is due to the fact that an optimum concentration of oxygen produces higher dissociation of oxygen molecules thus changes the ratio of atomic to molecular oxygen density. Shindo *et al.* also reported that the optimum value of α in microwave discharges in the argon-oxygen mixture at the oxygen concentration of about 15 percent [15]. Therefore, it is reasonable to expand the present study to include the mechanism of the production of negative ions theoretically as well as the effects of the axial magnetic field in detail in future studies.

## IV.  CONCLUSION

This work investigates the plasma density including negative ions, plasma potential and electron temperature in electronegative argon-oxygen discharge produced in a cylindrical capacitively coupled plasma discharge with axisymmetric magnetic field. The operating conditions such as applied RF power (25-75 watt), oxygen gas concentration (10-30%) and magnetic field strength (0-8 mT) are varied. The negative ion density is measured using the RF-compensated Langmuir probe from saturation current ratio method when a small quantity of oxygen gas is added to the pure argon discharge. The experimental results show an increase (~50%) and decrease (~40%) in the plasma density with a rise in RF power and oxygen concentration respectively. It is observed that the density of different species remains constant below 4 mT and thereafter decreases drastically. On the other hand, the electron temperature and plasma potential w.r.t. to floating shows opposite trend. This observed behaviour is due to the confinement of the high-energy electrons at the radial edge at a higher magnetic field since Larmor radius decrease below sheath thickness. Plasma potential is remains higher at low RF power in order to reduce the loss of electrons to the walls and to maintain the quasineutrality in the plasma bulk. In the lower magnetic field regime, i.e., below 4 mT, the gap between the positive ion density and the electron density is higher indicating a higher negative ion density. Whereas for the higher magnetic field regime, (above 4 mT), the gap between the positive ion and electron density decreases indicating a decrease in the negative ion density. Negative ion fraction is higher at lower magnetic fields and decreases with the increase in the magnetic field. We have observed a slightly higher negative ion fraction at 10 percent concentration of oxygen gas in the mixture. It shows that α is strongly dependent on the gas mixture and optimizing the composition of the gases is necessary for optimal negative ion fraction. Thus, in such discharge systems, the axial magnetic field acts as the controlling parameter to vary species density (positive ion, electron, and negative ions) and temperature as required for plasma processing applications. In conclusion, this experiment provides the ranges of magnetic field strengths for plasma confinement in cylindrical CCP discharges with axisymmetric magnetic field i.e. the field regime in which the density of the different species including negative ions is higher in such discharges.

# Acknowledgements

This work was supported by the Department of Atomic Energy, Government of India, and Science and Engineering Research Board (SERB), Core Research Grant No. CRG/2021/003536.